\pdfoutput=1
\documentclass[conference]{IEEEtran}
\makeatletter
\def\ps@headings{%
\def\@oddhead{\mbox{}\scriptsize\rightmark \hfil \thepage}%
\def\@evenhead{\scriptsize\thepage \hfil \leftmark\mbox{}}%
\def\@oddfoot{}%
\def\@evenfoot{}}
\makeatother
\usepackage[cmex10]{amsmath}
\usepackage{flushend,subfigure,booktabs,color,graphicx}
\newcommand{\argmax}{\operatornamewithlimits{arg\ max}}

\begin{document}
%
\title{Joint Ultra-wideband and Signal Strength-based Through-building Tracking for Tactical Operations}


\author{
\IEEEauthorblockN{Merrick McCracken, Maurizio Bocca, and Neal Patwari}
\IEEEauthorblockA{Electrical and Computer Engineering Department\\
University of Utah, Salt Lake City, Utah, USA\\
Email: merrick.mccracken@gmail.com, maurizio.bocca@utah.edu, npatwari@ece.utah.edu}
}


\maketitle

\thispagestyle{empty}

\begin{abstract}
  Accurate device free localization (DFL) based on received signal
  strength (RSS) measurements requires placement of radio transceivers
  on all sides of the target area.  Accuracy degrades dramatically if
  sensors do not surround the area.  However, law enforcement officers
  sometimes face situations where it is not possible or practical to
  place sensors on all sides of the target room or building. For
  example, for an armed subject barricaded in a motel room, police may
  be able to place sensors in adjacent rooms, but not in front of the
  room, where the subject would see them.  In this paper, we show that
  using two ultra-wideband (UWB) impulse radios, in addition to
  multiple RSS sensors, improves the localization accuracy,
  particularly on the axis where no sensors are placed (which we call
  the $x$-axis). We introduce three methods for combining the RSS and
  UWB data. By using UWB radios together with RSS sensors, it is still
  possible to localize a person through walls even when the devices
  are placed only on two sides of the target area. Including the data
  from the UWB radios can reduce the localization area of uncertainty
  by more than 60\%.
\end{abstract}

\begin{keywords}
Device-free localization, ultra-wideband, radio tomographic imaging, hidden Markov model, bistatic radar
\end{keywords}

\section{Introduction} \label{sec:introduction}

Device free localization (DFL) systems can be used in tactical
operations or crisis situations to help emergency personnel know where
people are in a room or building before they enter
\cite{Wilson_VRTI_2011}. These systems do not require people to
participate in the localization effort by wearing or carrying sensors
or radio devices. Radio frequency (RF) measurements are appropriate
for emergency operations because RF penetrates (non-metal) walls.
Such through-wall sensing could benefit police during hostage or
barricade situations.  However, in many such situations, it is not
possible to place sensors on all sides of the building or area.  For
example, some sides of a building might have windows where an armed
subject may be watching, and deploying sensors on that side could
expose police to harm or escalating the situation. As another example,
a room on an upper floor of a building may have some accessible
interior walls (\emph{e.g.}, in a hallway), but the exterior wall may
be unaccessible simply because of its height.  This paper provides a
system that expands the possibilities for RF-based DFL systems where
an area cannot be surrounded with sensors by combining RSS-based DFL
methods with bistatic ultra-wideband (UWB) impulse radar methods.

We are particularly motivated by discussions with our local SWAT team,
who have unfortunately, faced three situations in as many years in our
metro area \cite{abc4,sltrib,deseret} in which hostages were taken by
a barricaded subject in a hotel or motel room. Knowing the location of
the suspect represents very valuable information in planning the
actions (e.g., forced entry) required to bring the standoff to an end
safely.  In such situations, sensors could be placed in adjacent rooms
to the barricaded room, but rooms have front windows, and sometimes
back windows, thus front and back walls are potentially off-limits.

A DFL system based on received signal strength (RSS) measurements
\cite{youssef07,Wilson_RTI_2010,Patwari_2010_IEEE,Kaltiokallio_RTCSA_2011}
typically has radio transceivers, which we call \emph{RSS sensors}
placed on all four sides of a target area. Measured RSS between every
pair of sensors are used to estimate the location of the person in the
room in real-time. The localization process is based on models for the
change in RSS introduced by the presence of a person on or near the
\emph{link line}, \emph{i.e.}, the straight imaginary line connecting
the transmitter and receiver
\cite{Wilson_RTI_2010,Patwari_corr_shadow_2008,Wilson_SkewL_2011}. When
RSS sensors are placed only on two opposite sides of a room, the links
cross the monitored area along one axis but not the other. This
significantly degrades the localization accuracy of the system,
especially along the axis with no crossing links
\cite{Wilson_RTI_2010}.

UWB radios can be used for DFL through walls and can be accurate on
the order of centimeters or tens of cm \cite{Thoma, Bartoletti}.
Multiple UWB radios cooperating in a multistatic radar configuration
can provide an unambiguous localization estimate \cite{Thoma}.  A
transmitter broadcasts a UWB impulse and receivers capture the
time-domain channel impulse response (CIR) of the environment. Changes
to the CIR are detected and the time delay beyond the line-of-sight
(LoS) pulse for each of these changes is used to estimate the range of
the target from the radios \cite{mccracken2012hidden}. These radios,
however, can be prohibitively expensive to install on a permanent
basis: a single UWB impulse radio can cost thousands of dollars, and
using only a single pair of radios provides insufficient information
to unambiguously localize a target.

In this paper, we introduce a joint DFL system that uses the changes
measured in RSS and CIR to localize and track a target, such as a
person, through walls. We demonstrate, in particular, the localization
accuracy of a system which deploys sensors only on two opposite sides
of a room. We call the axis parallel to the sides of the room without
sensors the X axis and the axis parallel to the sides of the room with
sensors the Y axis (see Figures \ref{fig:study_room_layout} and
\ref{fig:hotel_layout}). The RSS sensors primarily provide the
information about the target's $y$ coordinate, while the UWB radios
primarily provide information about the target's $x$ coordinate. This
removes the need to have RSS sensors on all four sides of a target
room and reduces the number of UWB radios required for
localization. We introduce methods to process and combine the RSS and
CIR data in order to provide a unique position estimate. We also
propose methods that do not require an initial calibration of the
system. The experimental results collected in two deployments,
\emph{i.e.}, a study room at the University of Utah and a motel room
in Salt Lake City, show that the joint RSS-UWB DFL system can
accurately localize a non-cooperative target through walls. Even when
the number of deployed devices is low, \emph{e.g}, only two UWB radios
and six (three per side) RSS sensors, the system can still provide a
position estimate accurate enough to reliably indicate in which part
of the room the person is located. In tactical situations where the
only opportunity to have access to the target room is to open a breach
in a wall with an explosive frame, this information can be used by
police forces to decide which wall has to be detonated and avoid
hurting or killing the suspect.

In tactical operations or crisis situations, law enforcement may not
have the possibility of calibrating the systems used for DFL in
stationary conditions, \emph{i.e.}, when no person is located in the
target area. The methods used to process the data coming from the RSS
sensors and UWB radios should be able to localize and track the
suspect in the room from the start, making DFL a \emph{plug-and-play}
type of system. In this paper, we propose novel variance-based methods
for RSS and CIR measurements that can localize the person without
requiring an initial calibration of the system in stationary
conditions.

At the time of writing, there are several commercially available
through-wall radio technologies that can help law enforcement
determine the position of people inside a room. The Prism200 from
Cambridge Consultants \cite{Cambridge} is a through-wall radar system
for determining the location and movement of people for law
enforcement or emergency personnel. The Xaver\texttrademark\; products
from Camero are also through-wall UWB solutions that provide similar
capabilities \cite{Camero}. Time Domain is another company that offers
solutions for target localization and tracking using UWB radios
\cite{TD}. The UWB radios used in this work are a pair of P220 UWB
radios from Time Domain. Compared to these products, the joint RSS-UWB
DFL system described in this paper is considerably less expensive, as
the RSS sensors cost few tens of dollars each and only two UWB radios
are required. Moreover, the compact size and low weight of the RSS
sensors and UWB radios make our system easier to be installed.


The paper is organized as follows. In Section \ref{sec:rti}, we
describe the radio tomographic imaging (RTI) technique used to process
the RSS measurements coming from the RSS sensors. In Section
\ref{sec:uwb}, we describe the methods used for estimating the
bistatic range of a target using UWB radios by modeling the changes to
the CIR as a hidden Markov model. Section \ref{sec:combining}
introduces three methods to combine the RSS and CIR data in order to
provide a unique position estimate. Section \ref{sec:experiments}
describes the experiments carried out, while Section \ref{sec:results}
presents the results and compare the performance of the different
methods. Conclusions are given in Section \ref{sec:conclusions}.

\section{Radio Tomographic Imaging (RTI)} \label{sec:rti}

In RTI, originally introduced in \cite{Wilson_RTI_2010}, radio
transceivers placed at known positions form a wireless mesh network
and collect RSS measurements that can be used to localize and track a
person in real-time without requiring her to wear or carry any sensor
or radio device. RTI can provide sub-meter localization accuracy, also
in through-wall scenarios
\cite{Wilson_VRTI_2011,MASS,Grandma_2012}. The RSS measurements of all
the links of the network are processed in order to estimate a
discretized image $\mathbf{x}$ of the change in the propagation field
of the monitored area caused by the presence a person. The estimation
problem can be defined as:
\begin{equation} \label{RTI_linear_eq}
\mathbf{y} = \mathbf{W}\mathbf{x} + \mathbf{n},
\end{equation}
in which $\mathbf{y}$ and $\mathbf{n}$ are $L \times 1$ vectors of the
RSS measurements and noise of the $L$ links of the network,
respectively, and $\mathbf{x}$ is the $N \times 1$ image to be
estimated, where $N$ is the number of voxels of the image. Each
element $x_n$ of $\mathbf{x}$ represents the change in the propagation
field due to the presence of a person in voxel $n$. The $L \times N$
weight matrix $\mathbf{W}$ represents a spatial impact model between
the $L$ links of the network and the $N$ voxels of the image. The
model used in RTI
\cite{Wilson_VRTI_2011,Wilson_RTI_2010,Patwari_corr_shadow_2008,MASS}
is an ellipse having the foci located at the transmitter and receiver
of the the link. The voxels located within the ellipse have their
weight set to a constant which is inversely proportional to the root
distance between the transmitter and receiver, while the voxels
located outside of the ellipse have their weight set to zero.

\subsection{Attenuation-based RTI} \label{sec:attenuation_based_RTI}

For attenuation-based RTI, we use the method introduced in
\cite{MASS}. In this section, we briefly present this method and the
terminology that will be used also in the following sections.

The RSS of link $l$ on channel $c$ at time instant $k$, $r_{l,c}(k)$,
can be modeled as:
\begin{equation} \label{E:RSS}
 r_{l,c}(k) = P_{c} - L_{l,c} - S_{l,c}(k) + F_{l,c}(k) - \eta_{l,c}(k), \quad c \in \mathcal{F}
\end{equation}
where $P_{c}$ is the transmit power of the sensors, $L_{l,c}$ the
large scale path loss, $S_{l,c}$ the shadowing loss, $F_{l,c}$ the
fading gain (or \emph{fade level} \cite{Wilson_SkewL_2011}),
$\eta_{l,c}$ the measurement noise, and $\mathcal{F} = \{1, \ldots,
H\}$ is the set of measured frequency channels. Both the large scale
path loss $L_{l,c}$ and the shadowing loss $S_{l,c}$ change very
slowly with the center frequency. In our experiments, we use IEEE
802.15.4-compliant transceivers \cite{tidongle} which may transmit in
one of 16 channels across the $2.4$ GHz ISM band.  Because the band,
80 MHz, is small compared to 2.4 GHz, we can assume that both
$L_{l,c}$ and $S_{l,c}$ are independent of the frequency channel
$c$. Consequently, $F_{l,c}$ can be calculated as:
\begin{equation} \label{E:fade_level}
F_{l,c}(k) = r_{l,c}(k) - P_{c} + \eta_{l,c}(k).
\end{equation}
Due to the measurement noise $\eta_{l,c}$, the fade level can not be
measured directly. Thus, we estimate it by using the average RSS,
$\bar{r}_{l,c,}$, measured during an initial calibration of the system
performed when no person is in the monitored area:
\begin{equation} \label{E:fade_level_measure}
\bar{F}_{l,c} = \bar{r}_{l,c} - \min_{c}\bar{r}_{l,c}.
\end{equation}

In \cite{Wilson_SkewL_2011}, the links are divided in \emph{anti-fade}
and \emph{deep fade} links depending on the change in RSS measured
when a person crosses the \emph{link line}, \emph{i.e.} the imaginary
straight line connecting the transmitter and receiver. A link is in a
deeper fade on channel $c_1$ than on channel $c_2$ if $\bar{r}_{l,c_1}
< \bar{r}_{l,c_2}$. By definition, $\bar{F}_{l,c} \ge 0$ and
$\bar{F}_{l,c} = 0$ for one channel $c$ on each link. Anti-fade links
are the most informative for localization, since their spatial impact
area is limited around the link line, while deep fade links measure a
consistent change in RSS even when the person is far from the link
line. For this reason, for each link $l$ we calculate the fade level
in (\ref{E:fade_level_measure}) of each channel $c \in \mathcal{F}$,
and we rank the measured frequency channels from the most anti-fade to
the most deep fade. If $\mathcal{A}_i$ is the set of size $m$
containing the indices of the $m$ top channels in the fade-level
ranking, the link RSS measurement $y_l$ at time $k$ is calculated as:
\begin{equation}\label{E:MASS_RSS_change}
y_{l}(k) =  \frac{1}{m} \sum_{c \in \mathcal{A}_i} \Delta r_{l,c}(k),
\end{equation}
where $\Delta r_{l,c}(k) = r_{l,c}(k) - \bar{r}_{l,c}$, \emph{i.e.},
$\Delta r_{l,c}(k)$ is the difference between the current RSS
measurement and the average RSS measured during the initial
calibration phase.

\subsection{Variance-based RTI} \label{sec:variance_based_RTI}

We present a new multi-channel version of variance-based RTI extending
and improving the results of \cite{Wilson_VRTI_2011}. In this new
method, we also include the concept of fade level. We calculate the
mean of the RSS over a longer time window in order to better estimate
the fade level of each frequency channel, and we use this mean to
calculate the RSS short-term variance.

The attenuation-based RTI method in \cite{MASS} requires an initial
calibration of the system in stationary conditions, \emph{i.e.}, when
the monitored area is empty. Moreover, if the environment changes,
\emph{e.g.}, when the suspect in the room moves furniture or other
objects, the RTI system would need to be recalibrated or would
otherwise lose accuracy. The works in
\cite{Grandma_2012,Edelstein_2010} address this issue and introduce
methods capable of estimating the baseline RSS of the links on-line.

In tactical operations, such as when an armed person has barricaded
himself in a house or motel room before the arrival of police forces
on the scene, we cannot expect to require an empty
area. Variance-based RTI can be applied in this scenario. The change
in RSS due to the presence of a person on the link line can be
quantified as the unbiased sample variance of the last $N_s$ RSS
measurements:
\begin{equation}\label{E:short_term_variance_RSS}
\hat{s}_{l,c}(k) = \frac{1}{N_s-1} {\sum_{p=0}^{N_s-1}} \left( r_{l,c}(k-p)- \mu_{l,c}(k) \right)^2,
\end{equation}
where:
\begin{equation}\label{E:long_term_mean_RSS}
\mu_{l,c}(k) = \frac{1}{N_{\mu}} {\sum_{p=0}^{N_{\mu}-1}} r_{l,c}(k-p)
\end{equation}
is the mean of the last $N_{\mu}$ RSS measurements of link $l$ on
channel $c$, where $N_{\mu} > N_s$ so to estimate the mean RSS of each
channel on a longer time window and filter the changes due to the
person crossing the link line. Variance-based RTI does not require an
initial calibration of the system and can adapt at run-time to
eventual changes in the environment. For each link $l$, $\mu_{l,c}(k)$
in (\ref{E:long_term_mean_RSS}) provides an estimate of the fade level
of channel $c$ at time $k$. As for attenuation-based RTI in Section
\ref{sec:attenuation_based_RTI}, the channels are ranked from the most
anti-fade to the most deep fade. The link measurement $y_l$ at time
$k$ is calculated as:
\begin{equation}\label{E:MASS_RSS_change_var}
y_{l}(k) =  \frac{1}{m} \sum_{c \in \mathcal{A}_i} \hat{s}_{l,c}(k).
\end{equation}

\subsection{RTI image estimation} \label{sec:RTI_image_est} Since the
number of links $L$ is considerably smaller than the number of voxels
$N$, the estimation of the image $\mathbf{x}$ is an ill-posed inverse
problem that can be solved through regularization. In this work, we
use a regularized least-squares approach
\cite{Patwari_corr_shadow_2008,Zhao_SECON_2011,MASS,Grandma_2012}. The
discretized image of the change in the propagation field of the
monitored area is calculated as:
\begin{equation}\label{E:linear_transformation}
\hat{\mathbf{x}} =  \mathbf{\Pi}\mathbf{y},
\end{equation}
where $\mathbf{y}= [ y_1, \ldots, y_L]^T$, and
\begin{equation}\label{E:tikhonov}
{\mathbf{\Pi}} = {(\mathbf{W}^T\mathbf{W}+\mathbf{C}_{x}^{-1}\sigma_{N}^{2})}^{-1}\mathbf{W}^T,
\end{equation}
in which $\sigma_{N}$ is the regularization parameter. The \emph{a
  priori} covariance matrix $\mathbf{C}_{x}$ is calculated by using an
exponential spatial decay:
\begin{equation}\label{E:cov_matrix}
[\mathbf{C}_{x}]_{ji}=\sigma_{x}^{2}e^{-\| \mathbf{v}_{j}-\mathbf{v}_{i} \| /\delta_{c}},
\end{equation}
where $\sigma_{x}^{2}$ is the variance of voxel measurements, and
$\delta_{c}$ is the voxels' correlation distance. The linear
transformation $\mathbf{\Pi}$ is computed only once before the system
starts operating in real-time. The calculation of $\hat{\mathbf{x}}$
in (\ref{E:linear_transformation}) requires $L \times N$ operations
and can be performed in real-time. Table \ref{T:RTIImageParameters}
indicates the values of the parameters of the RTI image estimation
process.

\begin{table}[t!]
    \caption{RTI image estimation parameters} 
        \centering
        \footnotesize
        \begin{tabular}{l c c} 
        \hline\hline\          
        Description & Parameter & Value \\
        \hline  
        Voxel width [m]                    & $p$          & 0.15 \\
        Ellipse excess path length [m]     & $\lambda$    & 0.02 \\
        Voxels' variance [dB]              & $\sigma^2_{x}$ & 0.05 \\
        Noise standard deviation [dB]      & $\sigma_{N}$ & 1 \\
        Voxels' correlation distance       & $\delta_{c}$ & 4 \\
        Number of selected channels        & $m$          & 3 \\
        Short-term RSS variance window     & $N_s$        & 5 \\
        Long-term RSS mean window          & $N_{\mu}$    & 50 \\
        \hline  
        Empty area intensity threshold     & $T_e$        & 0.05 \\
        Number of updates for confirmation & $h_{app}$      & 8 \\
        Confirmation window                & $H$          & 15 \\
        Gating area radius [m]             & $\omega$     & 1.2  \\
        \hline 
        \end{tabular}
        \label{T:RTIImageParameters}
\end{table}

\section{Ultra-wide Band Range Estimation} \label{sec:uwb}

Assuming an UWB transmitter sends a pulse $\delta(t)$, each channel
impulse response (CIR) is measured as:
\begin{equation}
h(t) = \sum_i \alpha_i \delta(t-\tau_i),
\end{equation}
where $\alpha_i$ and $\tau_i$ are the complex amplitude and time delay
of the $i$th multipath component, respectively. The line of sight path
delay is $\tau_0$. The path delay of the target, which we wish to
estimate, is $\tau_*$. We will consider a discrete-sampled version of
the signal energy, $r_k$:
\begin{equation} \label{E:r_k}
r_k = \int_{(k-1/2)T}^{{(k+1/2)T}} |h(t)|^2 dt,
\end{equation}
where $T$ is the sampling period and $k$ ranges from $1 \ldots M$. In
this work, $T=1$\,ns. From now on, CIR time delays will be considered
only over discrete time intervals $k$ rather than continuously on $t$.

\subsection{Changes to the CIR as a Hidden Markov Model} \label{sec:uwb_hmm}

The changes to the UWB CIR are modeled as a hidden Markov chain. We
will refer to this method as \emph{HMM-UWB} or hidden Markov model
(HMM) based UWB. A hidden Markov chain is one whose states, $X_k=i$,
are not directly observable but are inferred from other observation
signals, $O_k$, available from the system. The distribution of the
observation signals is dependent on the state of the system,
\emph{i.e.}, $f_{O,i}=P(O|X=i)$. To estimate the probability the
system is in a given state at any time $k$, \emph{i.e.},
$P(X_k=i|\mathbf{O},\lambda)$, we need to know the distributions of
the observation signals, the initial state probabilities $\pi_i$, and
the state transition probabilities, $P_{i,j}$, all of which are
described by $\lambda = \left[ \pi_i, P_{ij}, f_{O,i} \right]$
\cite{Rabiner}.

The observations, $O_k$, are the difference between the CIRs of the
static environment and the CIRs of when a person is located in the
monitored area. This difference is calculated as the symmetric
Kullback-Leibler divergence, also known as relative entropy
\cite{Cover}. Assuming a Gaussian distribution for $r_k$ over many
CIRs, the closed form solution for the observed signal, $O_k$, is:
\begin{equation} \label{E:kld} O_k =
\frac{1}{2}\left(\frac{\sigma_p^2}{\sigma_q^2} +
\frac{\sigma_q^2}{\sigma_p^2} + \frac{\left(\mu_p-\mu_q\right)^2
\left(\sigma_p^2 + \sigma_q^2\right)}{ \sigma_p^2\sigma_q^2}\right) -
1
\end{equation} where $\mu_p$ and $\sigma_p^2$ are the mean and
variance of $r_k$ during the calibration period, and $\mu_q$ and $
\sigma_q^2$ are the mean and variance of $r_k$ when a person is
located in the target area, respectively. The distribution of the
observations is approximately a log-normal distribution \cite{Wisnet}.

If the changes to the CIR are modeled as a hidden Markov chain, the
CIR goes from an unchanged state, $X=0$, to a changed state, $X=1$, at
the time delay corresponding to the time traveled by the UWB pulse
from the transmitter to reflect off of the target and then arrive at
the receiving radio, \emph{i.e.} $k_*$, which is equivalent to
$\tau_*$. By applying this model to the system, standard HMM solving
algorithms, such as the forward-backward algorithm \cite{Rabiner}, can
be used to estimate when the system state changes and, thus, when
changes to the CIR occur. The forward-backward algorithm determines
the most likely state of the system at any given time as:
\begin{equation} \hat{X}_k = \argmax_i P(X_k=i|\mathbf{O},\lambda).
\end{equation} These state estimates are used to estimate $k_*$ as
\begin{equation} \label{E:k_*} 
P(X_k=1|\mathbf{O},\lambda) \leq 0.5 \} \hat{k}_* = \{ k_* | \hat{X}_k
\neq 1 \forall k < k_* \}
\end{equation} The work in \cite{McCracken} describes in further
detail the method for estimating UWB bistatic range and its improved
performance over other methods. From now on, we will let $\alpha_k =
P(X_k=1|\mathbf{O},\lambda)$. $\alpha_k$ describes the probability
those CIRs possibly affected by a person at time $k$ are in the
affected state. These probabilities are used to form the UWB
localization image.

When solving the forward-backward algorithm, accurate estimates of
when state changes occur is dependent on how well $\lambda$ models the
true system parameters. Often $\lambda$, and particularly the
distribution of observation signals, is unknown when beginning to
solve the forward-backward algorithm. A known $\lambda$ from another
environment can be used as an initial estimate for $\lambda$ when
solving for the state estimates. With more observations and state
estimates as time progresses, the Baum-Welch algorithm can help tune
$\lambda$ to more closely match the true parameters and improve range
estimates \cite{Rabiner, McCracken}.

It is critical that the observation vectors are accurate
representations of the difference between a CIR with a possible target
and the CIR of the known environment. This is a function of how well
the calibration CIRs represent the static environment. Even as a
person is walking in a room, something in the environment may change
that can significantly affect the accuracy of the range estimates due
to a change in the static environment. This includes changing the
position of furniture or a door opening or closing. If the bistatic
range of the change to the environment is closer than the target, it
will not be possible to estimate the range of the target using the
existing calibration CIRs. Such a change to the environment requires
that the calibration CIRs be updated to reflect this change, which is
very difficult with a person moving within the target
environment. This is a primary disadvantage of this hidden Markov
model for detecting changes to the CIR.

The most extreme case of choosing calibration CIRs would be to use the
CIR samples that immediately precede the samples with a possible
target. These calibration CIRs may include the target of interest as
part of what is considered the static environment. This would minimize
environmental variation that may occur over time and effectively
eliminate the calibration requirement. This would also have a negative
effect of biasing the target's range estimate toward the radios if the
target is moving away from the radios and makes the detection of a
non-moving target more difficult. However, if CIRs are sampled
frequently enough relative to the person's speed, this bias toward the
radios would be minimal.

In this work, we assume there are no major changes to the environment
throughout each trial that would require new calibration CIRs to be
captured. This allows us to use just one calibration period for
estimating $k_*$.

\subsection{Variance-based UWB Range Estimation} \label{sec:uwb_variance}
 
An alternative method is to use the short-term variance of the CIR for
each $r_k$. We refer to this method as \emph{VB-UWB}, or
variance-based UWB. $\alpha_k$ is calculated as:
\begin{equation}
\alpha_k = \frac{\sigma^2_{r_k}}{g_{r_k}},
\end{equation}
where the variance $\sigma^2_{r_k}$ is the unbiased sample variance of
$r_k$ over the $N_U$ most recent CIRs. In this work, we let $N_U=5$,
corresponding to the number of CIRs captured in approximately $0.5$
s. The normalization coefficient $g$ is calculated as:
\begin{equation}
g = g(1-\beta) + r_k\beta.
\end{equation}
This is equivalent to applying a low-pass IIR filter to $r_k$. In this
work, $\beta = \frac{1}{N}$. Because the variance of $r_k$ is high
when the mean of $r_k$ is high and vice versa, we normalize the
variance $\sigma^2_{r_k}$ by the mean of $r_k$. In this way,
$\alpha_k$ increases only when the person moves. This method is used
in conjunction with the variance-based RTI method described in Section
\ref{sec:attenuation_based_RTI}. The primary advantage of this method
is that no calibration is required to solve for $\alpha_k$. A
disadvantage is that the target can disappear if it remains motionless
over a long period of time. We alleviate this problem by applying the
methods in Section \ref{sec:gating}.

\subsection{UWB Image Estimation} \label{sec:uwb_image}

When estimating the UWB image, the image space is constrained to
contain only the inner dimensions of the target room plus one
additional voxel on each image edge. Discretizing the image space into
$N$ voxels, the image vector is:
\begin{equation}
  \mathbf{l}^u = [l^u_1, \ldots, l^u_N]^T,
\end{equation}
where each voxel $l^u_n$ has a bistatic range to the UWB transmitter
and receiver described by its path delay $k_n$. The value of each
voxel, $l^u_n$, is calculated as the non-negative difference function:
\begin{equation}
l^u_n = (\alpha_{k_n} - \alpha_{k_n -1})^{+}
\end{equation}
where the non-negative difference function is defined as:
\begin{equation}
(x)^{+} = 
   \begin{cases}
     x & \text{if } x \geq 0 \\
     0 & \text{if } x < 0
   \end{cases},
\end{equation}
and assuming $\alpha_0 = 0$.

\begin{table}[t!]
    \caption{UWB estimation parameters} 
        \centering
        \footnotesize
        \begin{tabular}{l c c} 
        \hline\hline\          
        Description & Parameter & Value \\
        \hline  
        Voxel width [m]                    & $p$          & 0.15  \\
        Sampling Period [ns]               & $T$          & 1 \\
        Short-term CIR variance window     & $N_U$        & 5 \\
        Variance normalization parameter   & $\beta$      & 1/$N_U$ \\
        \hline 
        \end{tabular}
        \label{T:UWBImageParameters}
\end{table}

\section{Combining RTI and UWB Information} \label{sec:combining} In
this section we introduce three methods to combine RTI and UWB and we
compare the results of the different methods in Section
\ref{sec:results}.

\subsection{Image Combination} \label{sec:image_product}

An RTI image is formed as described in Section
\ref{sec:attenuation_based_RTI} after every RSS sensor has transmitted
a packet on all channels in $\mathcal{F}$, \emph{i.e.}, after RSS
measurements have been collected on all the links and channels. A UWB
image is formed for every new CIR captured. The two images are
combined to form the new image $L^c$ by performing a voxel-wise
product,
\begin{equation}
L^c = l^r \wedge l^u.
\end{equation}
where $l^r=\hat{\mathbf{x}}$ from Equation
(\ref{E:linear_transformation}) and $l^u$ is from the UWB image
$\mathbf{l}^u$. We define $M_{L^c} = \max{L^c}$. When no person is
located in the monitored area, $M_{L^c}$ has a very low value. We use
a threshold $T_e$ to avoid further processing images not showing the
presence of a person in the target area: if $M_{L^c} \le T_e$, we
discard the current combined image and wait for the next one formed by
the system. Otherwise, we normalize the values of the voxels of $l^r$
and $l^u$ such that their minimum value is zero and the sum of all
voxels is one:
\begin{equation}
[\hat{l}^r]_{n} = \frac{{{l}^r}_{n}}{\sum_{i=1}^N {{l}^r}_{i}}, 
\end{equation}
and similarly for $l^u$:
\begin{equation}
[\hat{l}^u]_{n} = \frac{{{l}^u}_{n}}{\sum_{i=1}^N {{l}^u}_{i}}.
\end{equation}
The normalization brings the two images in the same range of values
and weights them equally. The normalized combined image $\hat{L}^c$ is
calculated again by performing a voxel-wise product of $\hat{l}^r$ and
$\hat{l}^u$:
\begin{equation}
\hat{L}^c = \hat{l}^r \wedge \hat{l}^u.
\end{equation}
The RSS and UWB data collected by the two systems are time stamped to
allow synchronizing the two images. Images are formed at the same rate
as the higher of the two sampling rates. In our case, since the UWB
CIRs are sampled more frequently than each RTI cycle, a combined image
is formed for each new UWB sample. This image will then be the
combination of the most recently formed RTI with the new UWB
image. From the normalized combined image, $\hat{L}^c$, the position
of the person is estimated as:
\begin{equation} \label{E:person_position}
\hat{p} = \arg\max_{n \in N} \hat{L}^c,
\end{equation}
\emph{i.e.}, the person's position estimate is at the voxel $n$ having the highest value.

\subsection{Linear Inversion with UWB Data} \label{sec:uwb_inversion}

An alternative method to form a combined localization image is to
modify the weight $W$ matrix in (\ref{RTI_linear_eq}) to include the
UWB measurements in the inversion process. We define the matrix
$\mathbf{W}_U$ is an $M \times N$ matrix where $M$ is the maximum
value of $k$ and $N$ is the number of voxels of the image. Each column
of $\mathbf{W}_U$ represents the ideal vector of $\alpha_k$ if the
target were located at voxel $p$. The vector $y_U$ is the estimated
vector of $\alpha_k$ from the results of the forward-backward
algorithm. Equation (\ref{RTI_linear_eq}) then becomes:
\begin{equation}
    \begin{bmatrix}
        \mathbf{y}_R \\
        \mathbf{y}_U
    \end{bmatrix}
    =
    \begin{bmatrix}
        \mathbf{W}_R \\
        \mathbf{W}_U
    \end{bmatrix} \mathbf{x}
    +
    \begin{bmatrix}
        \mathbf{n}_R \\
        \mathbf{n}_U
    \end{bmatrix}
\end{equation}
where the subscripts $R$ and $U$ correspond to the matrices derived
from the RSS or UWB data, respectively. The inversion matrix is
calculated as described in Section \ref{sec:RTI_image_est} using the
combined $W^C$ matrix. A combined localization image $\hat{L}^c$ is
then formed by multiplying the inversion matrix in (\ref{E:tikhonov})
to the combined RSS and UWB measurement matrix $\mathbf{y}^c$. The
position of the person is estimated as in (\ref{E:person_position}).

\subsection{Estimating X by using Y} \label{sec:x_from_y}

Another method for combining the UWB and RTI images is to derive one
coordinate of the position estimate of a target from each
image. First, we estimate the target location from the RTI image
formed as described in Section \ref{sec:attenuation_based_RTI}. The
\emph{y}-coordinate from this position estimate is used to derive an
\emph{x}-coordinate from the UWB image, which is calculated as
described in \ref{sec:uwb_image}. If the target location estimate from
the RTI image is at coordinates $(\hat{x}^R,\hat{y}^R)$, we consider
the row of the UWB image corresponding to $\hat{y}^R$. The target
position estimate $\hat{p}$ is set at the voxel having the maximum
value in that row, \emph{i.e.}, $\hat{p} = (\hat{x}^U,\hat{y}^R)$.

\section{Localization and Tracking} \label{sec:gating}

The position estimate $\hat{p}$ is used for updating an already
existing track of a person or for initiating a new one if the target
area is empty. To this purpose, we use track confirmation and deletion
rules \cite{blackman99}. If at time $k$ the set of \emph{candidate
  tracks}, $\mathcal{T}_d$, and the set of \emph{confirmed tracks}
$\mathcal{T}_f$ are both empty, \emph{i.e.}, the position estimate
$\hat{p}(k)$ is used to start a new candidate track, which is added to
$\mathcal{T}_d$. A candidate track becomes a confirmed track only if
its position has been updated at least $h_{app}$ times in the last $H$
formed combined images ($h_{app} \le H$). If this condition is not
fulfilled, the candidate track is deleted.

A \emph{gating area} of radius $\omega$ is centered at the target's
position estimate $\hat{p}$. The radius $\omega$ is defined as an
integer multiple of the voxel width $p$. If one or both of the sets
$\mathcal{T}_f$ and $\mathcal{T}_c$ are not empty, only the tracks
located within the gating area are considered for a potential
association to the current position estimate. These tracks form the
set $\mathcal{T}_g \subseteq (\mathcal{T}_f \cup \mathcal{T}_c)$. The
confirmed tracks in $\mathcal{T}_g$ are given priority over the
candidate tracks: the current position estimate is used to update the
closest confirmed track. Otherwise, if no confirmed track exists, the
current position estimate is used to update the closest candidate
track. If the set $\mathcal{T}_g$ is empty, the current position
estimate is used to start a new candidate track. By using the gating
area, we avoid the position estimate of the person to have large
sudden changes in correspondence of noisy RSS and CIR measurements
from the two systems.

\section{Experiments} \label{sec:experiments}

\begin{figure}[t!]
    \centerline{\includegraphics{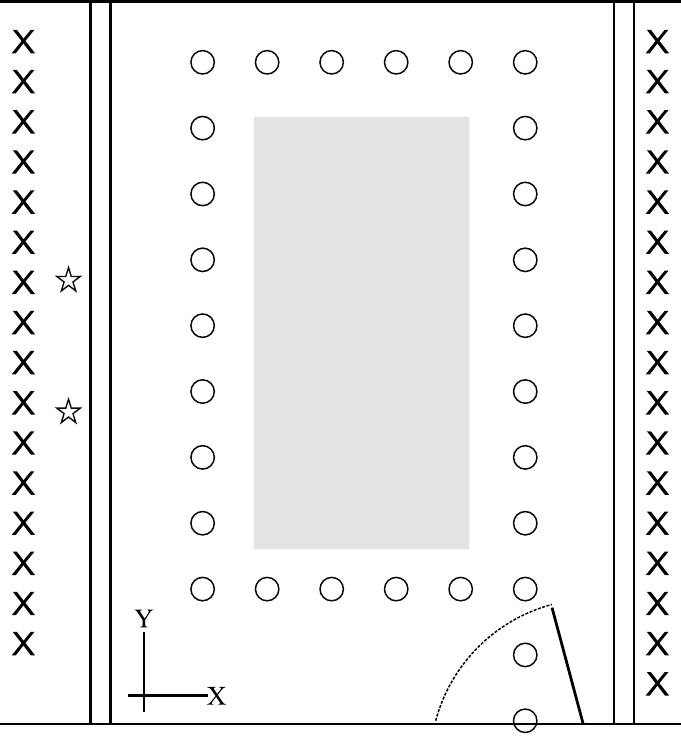}}
    \caption{Layout of the study room located in the Warnock
      Engineering Building at the University of Utah used for the
      experiments. Xs represent the $33$ RSS sensors. Stars represent
      the $2$ UWB radios. Circles represent the steps taken by the
      person at one second intervals. Grey rectangles represent
      furniture. The target room's inner dimensions are $3.82$ m by
      $5.49$ m ($21$ m$^{2}$ area).}
    \label{fig:study_room_layout}
\end{figure}

The first experiment was conducted in a study room on the second floor
of the Warnock Engineering Building at the University of Utah. A total
of $33$ RSS sensors were placed outside of the room along two opposite
walls, $17$ on one side and $16$ on the other. The sensors were $30.5$
cm apart. Two UWB radios were placed on one of the two sides of the
room where the RSS sensors were positioned. The UWB radios were $1$ m
apart. A person walked along a predefined path six times, three times
counter-clockwise and three times clockwise. The person entered and
exited the room in each of the six trials. With the help of a
metronome and markings on the floor, the person walked at a constant
speed of $0.5$ m/s. Figure \ref{fig:study_room_layout} shows the setup
of the tests carried out in the study room.

\begin{figure}[t!]
    \centerline{\includegraphics{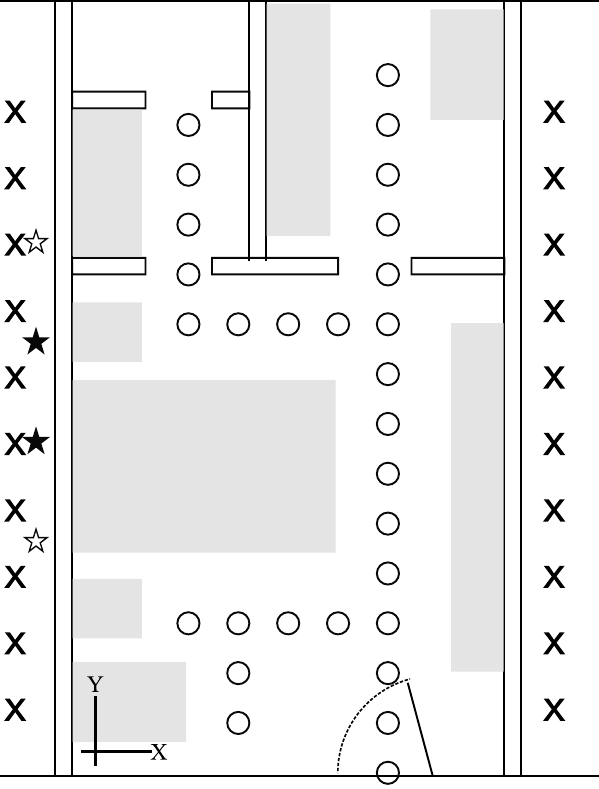}}
    \caption{Layout of the room of a motel located in Salt Lake City,
      Utah. Xs represent the RSS sensors. White and black stars
      represent the UWB radios in configurations A and B,
      respectively. Circles represent the steps taken by the person at
      one second intervals. Grey rectangles represent furniture. The
      target room's inner dimensions are 3.96 m by 7.11 m ($28$
      m$^{2}$ area).}
    \label{fig:hotel_layout}
\end{figure}

\begin{figure}[t!]
    \centerline{\includegraphics[width=2.7in]{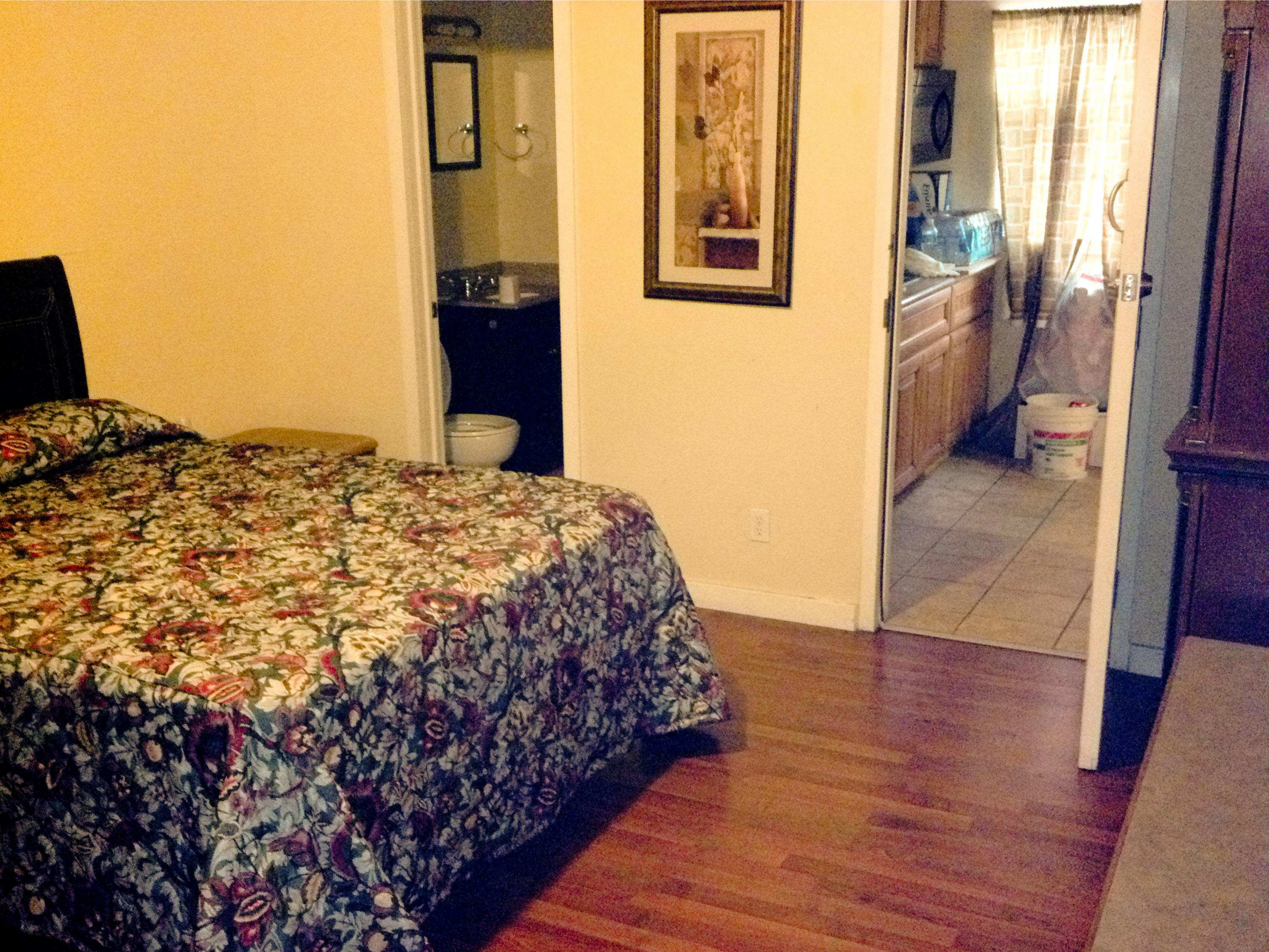}}
    \caption{Photo of the interior of the motel room where the second
      experiment was conducted.}
    \label{fig:inside_hotel}
\end{figure}

The second experiment was conducted in a room of a motel in Salt Lake
City, Utah. The layout of this room is described in Figure
\ref{fig:hotel_layout}. This time, ten RSS sensors were placed along
each of the walls separating the room from the adjacent ones. Two UWB
radios were placed outside one wall of the target room. The
experiments were conducted with the UWB radios at two different
distances, $0.9$ m and $2.7$ m apart. A person walked along a
predefined path at a constant speed of $0.5$ m/s, entering and exiting
the room each trial. There were no other rooms adjacent to the target
room besides the two where sensors were placed.  For the second
experiment, a person walked the target path $18$ times. Six of the
trials were done with the UWB radios in configuration A and twelve in
configuration B, represented by white stars and black stars,
respectively, in Figure \ref{fig:hotel_layout}. A photo of the
interior of the target room can be seen in Figure
\ref{fig:inside_hotel}.

\section{Results} \label{sec:results}

For the first experiment, 50 simulations were run using randomly
selected subsets of $S$ RSS sensors available on each side of the
room. The density of sensors on each side of the target room is higher
than what would be used in a typical deployment. Subset sizes for
these simulations ranged from 3 to 10 sensors per side. The same
number of sensors were used on each side of the room. The same subset
of sensors was used for each of the six trials and remained the same
when UWB radio data was included for a given simulation. The gating
algorithm described in Section \ref{sec:gating} was applied in all
simulations. Simulations were performed using AB-RTI, AB-RTI with
HMM-UWB, VB-RTI, and VB-RTI with VB-UWB.

\begin{figure*}[htb]
    \centerline{\includegraphics{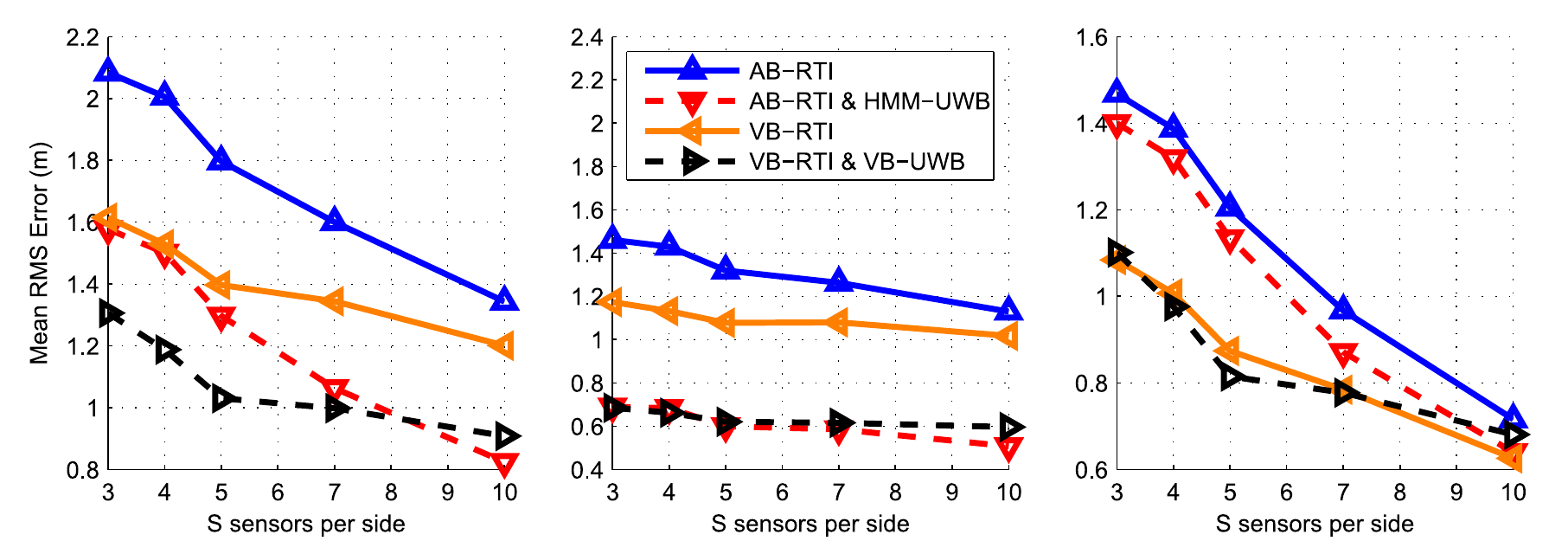}}
    \caption{From left to right, the mean RMS L2, X, and Y errors over the 6 trials and 50 simulations using random subsets of $S$ sensors per side of the study room.}
    \label{fig:exp2_allerr}
\end{figure*}

Figure \ref{fig:exp2_allerr} shows the mean RMS localization error for
each of the methods used measured. Each point on the figure is the
error is averaged over the 50 simulations and 6 trials performed for
this experiment using $S$ sensors.

The Y-axis error improves significantly with each additional sensor
used on each side of the room. There is also little improvement in the
Y-axis error as a result of including the UWB information. This is the
expected behavior of the system. Variance-based methods show
improvement in reducing Y error over attenuation-based methods.  The
X-axis error improves as a result of including more sensors on each
side of the room but not as greatly as does the Y-axis error. The
improvement as a result of including UWB information, however, is much
more significant and is also almost constant as a function of the
number of RSS sensors. The localization error improves overall, that
is, as a Euclidean distance, by 51 cm and 33 cm, on average, for
attenuation and variance-based methods, respectively.

As a point of comparison, if a point in the room is selected at random
at each time, the RMS L2 error is 2.94 m on average over the 6
trials. Errors for the X and Y axes by selecting random locations are
1.65 m and 2.44 m, respectively. The gating algorithm is not applied
when performing estimating location using random coordinates.

\begin{figure*}[htb]
    \centerline{\includegraphics{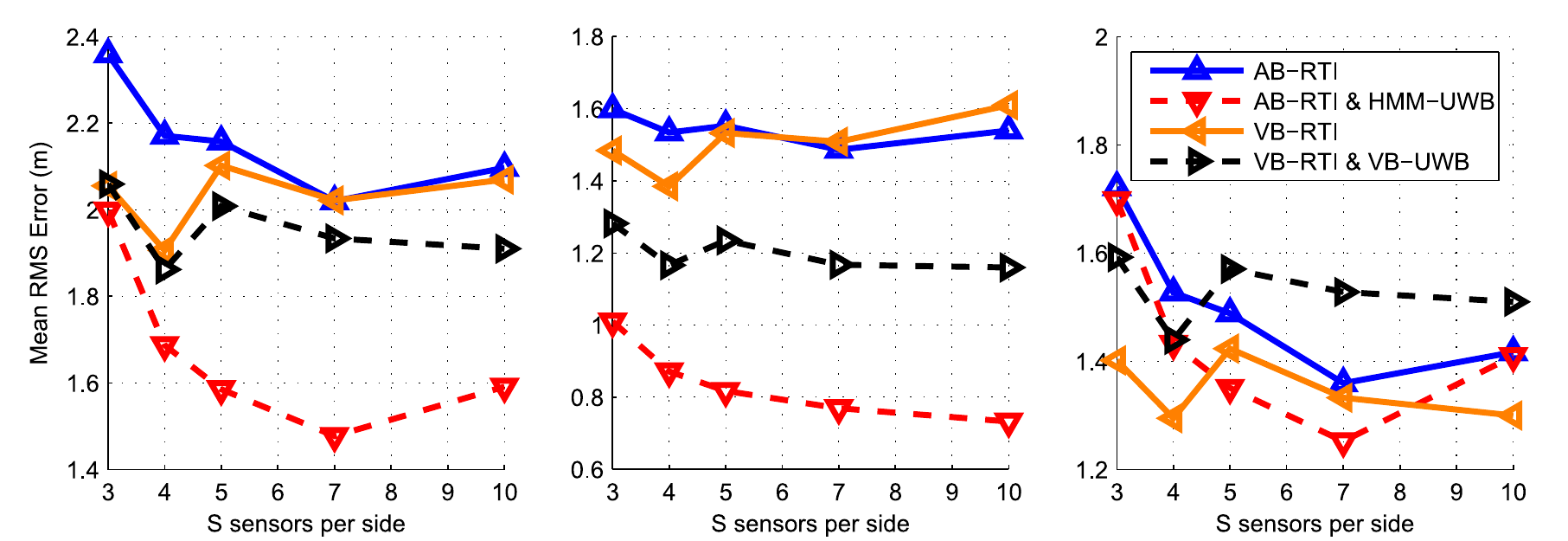}}
    \caption{From left to right, the mean RMS L2, X, and Y errors over
      the 18 trials and 50 simulations using random subsets of $S$
      sensors per side of the motel room. When $S=10$, only 1
      simulation was performed.}
    \label{fig:hotel3_allerr}
\end{figure*}

For the second experiment, 50 simulations were also run using randomly
selected subsets of $S$ RSS sensors on each side of the room for each
simulation. When $S=10$, however, only one simulation was performed
because there was only one possible combination of $S=10$ radios per
side. For each simulation, localization is performed using AB-RTI,
AB-RTI with HMM-UWB, VB-RTI, and VB-RTI with VB-UWB.
The gating algorithm described in Section \ref{sec:gating} is also
applied to each of these methods. Figures \ref{fig:hotel3_allerr}
shows, from left to right, the L2, X, and Y errors when applying these
four methods to the data collected over the 18 trials performed in the
motel room.

One noticeable difference between the results of the two experiments
is that the Y error in the second experiment degrades significantly by
including VB-UWB with VB-RTI whereas for the first experiment the Y
error was effectively the same. Generally, however, the same trends
are visible in the results for the second experiment. The Y error
improves with increasing $S$ and including UWB data significantly
improves X error.

For the second experiment, the error using 10 sensors per side is
higher than the error using 7 sensors, in many cases. There were only
10 sensors on each side of the room and, therefore, only one unique
simulation could be performed using 10 sensors. By performing many
simulations using subsets of the available sensors, the effects due to
sensor placement on localization error could be minimized. This was
not possible in the case where $S=10$ for the second experiment.

Table \ref{tab:experiment2_results_table} shows the mean RMS error
over the 18 trials performed for this experiment using all 20 RSS
sensors. For comparison and as an estimate of the upper bound on error
for a given environment and target path, random image coordinates are
selected as the target location estimate. At each time when a combined
image would be formed, X and Y coordinates and randomly selected and
are used as the location estimate at that time. The gating algorithm
described in Section \ref{sec:gating} is not applied when randomly
choosing location estimates. The results from applying the methods
described in Sections \ref{sec:uwb_inversion} and \ref{sec:x_from_y}
are also given in Table \ref{tab:experiment2_results_table}.

\begin{table*}[htbp]
  \centering
  \caption{Mean RMS localization error for the second experiment over all 18 trials for the methods described. Gating was used for all methods except random selection. Units given in meters.}
    \begin{tabular}{lccccccc}
    \toprule
          & AB-RTI & AB-RTI \& HMM-UWB & VB-RTI & VB-RTI \& VB-UWB & Inversion with UWB & X from Y & Random \\
    \midrule
    L2    & 2.10  & 1.59  & 2.07  & 1.91  & 1.84  & 1.76  & 3.31 \\
    X     & 1.54  & 0.73  & 1.61  & 1.16  & 1.31  & 0.98  & 1.53 \\
    Y     & 1.42  & 1.41  & 1.30  & 1.51  & 1.28  & 1.44  & 2.94 \\
    \bottomrule
    \end{tabular}%
  \label{tab:experiment2_results_table}%
\end{table*}%


Note in Table \ref{tab:experiment2_results_table} that when performing
localization using AB-RTI or VB-RTI, the X-axis error is about the
same as that obtained from randomly guessing an X coordinate for each
image. This is critically important for tactical operations. Having
some knowledge about the person's coordinate in each axis is essential
for law enforcement personnel to be able to make tactical decisions.
 


\subsection{Area of Uncertainty}

We define the area of uncertainty (AoU) as the ratio of the L2 mean
squared error (MSE) to the total area of the monitored room:
\begin{equation}
AoU = \frac{\textrm{L2 MSE}}{\textrm{Room Area}}.
\end{equation}
Table \ref{tab:aou_reduction} shows the percent reduction in the AoU
by adding UWB data to AB-RTI and VB-RTI for $S=3$ and $S=10$ sensors.

\begin{table}[bthp]
  \centering
  \caption{Percent reduction of AoU by including UWB data.}
    \begin{tabular}{lccccc}
    \toprule
          & \multicolumn{2}{c}{Study Room} & & \multicolumn{2}{c}{Motel Room} \\
          & AB-RTI & VB-RTI & & AB-RTI & VB-RTI \\
    \cline{2-3} \cline{5-6}
    $S=3$   & 40.2\% & 32.4\% & & 26.3\% & 0.2\% \\
    $S=10$  & 61.8\% & 43.2\% & & 41.3\% & 14.9\% \\
    \bottomrule
    \end{tabular}%
  \label{tab:aou_reduction}%
\end{table}%

The percent reduction in the AoU is significant except for VB-RTI in
the motel room using $3$ sensors. This may be due to the particular
subsets of sensors used in the simulations when $S=3$. The reduction
in the AoU confirms that by adding UWB data the system can more
accurately indicate to law enforcement personnel in which part of the
room the person is located.

\section{Conclusions} \label{sec:conclusions}

In this work, we present a joint DFL system that uses the changes
measured in RSS and UWB CIR to localize and track a person through
walls. We target tactical operations and crisis situations where it is
not possible for the police forces to place sensors on all sides of
the area to be monitored. Experimental results show that including UWB
with RSS data significantly improves localization accuracy when RSS
sensors are only available on two sides of the target area. Where RSS
sensors have been placed along the Y axis, improvements in accuracy
along the X axis by including UWB data are especially
significant. Without including UWB data, the accuracy along the X axis
can be as bad as randomly guessing an X coordinate.

We introduce three methods to combine the information from the UWB and
RSS systems and we compare their performance. The multi-channel
variance-based RTI method proposed in this work, which does not
require an initial calibration in stationary conditions, is as
effective or more effective than attenuation-based RTI for
through-wall localization. The improvements in localization accuracy
and the reduction in the AoU demonstrate that UWB data should be
included in a DFL system for tactical operations where RSS sensors
may only be placed on two sides of a room.

\bibliographystyle{IEEEtran}
\bibliography{IEEEabrv,uwb-rti} 

\end{document}